\documentclass[notitlepage, twoside,english,nofootinbib,preprintnumbers,aps,11pt,showpacs]{revtex4-1} \usepackage[a4paper, margin=1.5in]{geometry} \usepackage{amssymb} \usepackage{commath}
\usepackage{amsmath}
\usepackage{url}
\usepackage{float}
 \usepackage[utf8]{inputenc} \usepackage[T1]{fontenc} \usepackage{lmodern}
\usepackage{braket}
\usepackage{graphicx}
\usepackage{pictexwd,dcpic}
\usepackage{braket}
\usepackage{atbegshi,picture, afterpage}
\usepackage{lipsum}
\newcommand{\be}{\begin{equation}}
\newcommand{\ee}{\end{equation}}
\newcommand{\ba}{\begin{align}}
\newcommand{\ea}{\end{align}}
\begin{document}
\title{Extremal horizons stationary to the second order: \\ new constraints}
\author{Maciej Kolanowski}
\email[]{mp.kolanowski@student.uw.edu.pl}
\author{Jerzy Lewandowski}
\email[]{Jerzy.Lewandowski@fuw.edu.pl}
\author{Adam Szereszewski}
\email[]{Adam.Szereszewski@fuw.edu.pl}
\affiliation{\vspace{6pt} Institute of Theoretical Physics, Faculty of
  Physics, University of Warsaw, Pasteura 5, 02-093 Warsaw, Poland}
  \date{\today}
\begin{abstract}
   {We consider non-expanding shear free (NE-SF) null surface geometries embeddable as extremal Killing horizons to the second order in Einstein vacuum spacetimes. A NE-SF null surface geometry consists of a degenerate metric tensor and a consistent torsion free covariant derivative. We derive the constraints implied by the existence of an embedding. The first constraint is well known as the near horizon geometry equation. The second constraint we find is new. The constraints lead to a complete characterization of those NE-SF null geometries that are embeddable in the extremal Kerr spacetime. Our results are also valid for spacetimes with a cosmological constant.}
\end{abstract}
\maketitle
\section{Introduction}
Extremal horizons attracted a lot of attention in the past years. They allowed for a microscopic, string-theoretic derivation of Bekenstein--Hawking formula within gauge/gravity correspondence framework \cite{Strominger:1997eq} which was further generalized to Kerr/CFT correspondence \cite{Guica:2008mu}. They were also a missing piece of uniqueness theorems for black holes \cite{Amsel:2009et, Chrusciel:2012jk}. Furthermore, extremal horizons gave rise to the new class of solutions to Einstein equations -- near-horizon geometries \cite{Bardeen:1999px, Pawlowski:2003ys, Lewandowski:2016sou}. \\
The goal of our work is derivation of  constraints that are satisfied by extremal horizon's geometry. The horizon geometry consists of induced metric tensor  and covariant derivative. It is equivalent to the characteristic Cauchy data defined on the horizon. The first constraint follows from the Killing equation satisfied at the horizon to the fist order and is well known in the literature  as the Near Horizon Geometry equation. A metric $g_{AB}$ induced on a spatial section $\mathcal{S}$ of the horizon embedded in $n$-dimensional spacetime (satisfying Einstein equations with a cosmological constant $\Lambda$ of arbitrary sign) must satisfy the following equation \cite{Hajicek:1974oua, Moncrief:1983xua, Ashtekar:2001jb, Lewandowski:2004sh}:
\begin{equation}
    0 = \nabla_{(A} \omega_{B)} + \omega_A \omega_B - \frac{1}{2} \stackrel{(n-2)}{R_{AB}} + \frac{\Lambda}{n-2} g_{AB}, \label{nhge}
\end{equation}
where $\omega_A$ is one-form on $\mathcal{S}$, $\nabla_A$ is Levi-Cevita connection of $g_{AB}$ and $\stackrel{(n-2)}{R_{AB}}$ its Ricci tensor. Despite various attempts and many results (for example, \cite{Lewandowski:2002ua, Kunduri:2008rs, Chrusciel:2017vie, Chrusciel:2005pa, Dobkowski-Rylko:2018nti}, see \cite{Kunduri:2013ana} and references therein for a more systematic discussion)  existence and  uniqueness of solutions to \eqref{nhge} is still an open question. Still, however, part of the data defining the extremal horizon geometry remains unconstrained.     In this work we go a step further and, assuming the Killing horizon equation to the second order. 
We managed to derive constraint equations on  other elements of the horizon geometry.  \\
The rest of this paper is organized as follows. We start with the basic definitions of horizon structures in section \ref{INS}. Then, in Section III we derive the new constraints in a simple manner 
using the Newman–Penrose formalism available in $4$ spacetime 
dimensions. In  Section IV, we generalize our result to arbitrary 
dimension. The technical details are relegated to the Appendix. We discuss our result, compare it with the literature and analyze its possible significance in section \ref{dis}.

\section{Isolated null surfaces} \label{INS}
In this paper we consider  an $n$-dimensional  spacetime that consist of a manifold  $M$  and a metric tensor $g_{\mu\nu}$ of the signature $-+...+$.  By $\nabla_\mu$  denote the  torsion free  covariant derivative in $M$,  corresponding to $g_{\mu\nu}$ via
$$\nabla_\alpha g_{\mu\nu}\ =\ 0.$$   
We assume  the vacuum Einstein equations with a cosmological constant $\Lambda$,
\begin{equation}\label{EE} G_{\mu\nu} + \Lambda g_{\mu\nu}\ =\ 0, \end{equation}
where $G_{\mu\nu}$ is the Einstein tensor.  

\subsection{Notation and convention}\label{INSA}
Throughout this paper we use the following (abstract) index  notation (\cite{Wald}):
\begin{itemize}
\item  Indices of the spacetime tensors  are denoted by
lower Greek letters: $ \alpha, \beta,\gamma, ... \ = 1,2,...,n,$
\item  Tensors defined in $n-1$-dimensional  surfaces $H\subset M$  carry indices denoted by lower Latin letters: $a, b, c, ...\ = 1,...,n-1$
\item  Capital Latin letters $A, B, C, ... \ = 1,...,n-2$ are used as the indices of tensors defined in an $(n-2)$-dimensional sub-surfaces $S\subset H$.  \end{itemize}
\subsection{Non-expanding shear-free null surfaces}\label{NE-SF} 
In $M$ consider an $n-1$ dimensional null surface $H$. 
The spacetime metric tensor $g_{\mu\nu}$ induces in $H$ a degenerate metric tensor $g_{ab}$. 
The degeneracy means, that at every point  $x\in H$ there is a vector $0\not=\ell \in T_xH$
such that 
\begin{equation}\label{lh}
\ell^a g_{ab}\ =\ 0 .
\end{equation}
In other words, $H$ is orthogonal to $\ell$. In particular
\be\label{ll} \ell^\mu \ell_\mu =0.\ee
The integral curves of the distribution of the null directions  locally foliate $H$ and each of the curves is  geodesic in the spacetime $ M$.  
We refer to them as null generators of $H$.  
\medskip

We call a null surface $H$ non-expanding and shear-free (NE-SF) if  for every pair $X$ and $Y$ of vector fields tangent to $H$,  the spacetime 
vector field  $X^\alpha\nabla_\alpha Y^\mu$ is also tangent to $H$,
\begin{equation}
X,Y\in \Gamma(T(H)) \ \ \Rightarrow \ \ \nabla_X Y \in \Gamma(T(H)) .  
\end{equation}
In other words,  the spacetime covariant derivative $\nabla_\mu$  preserves the tangent bundle 
$T(H)$, a sub-bundle of $T(M)$, and via the restriction induces on $H$  a  covariant derivative $\nabla_a$ . 
\medskip 

\noindent{\bf Definition 1} {\it Given a null NE-SF surface $H$, the pair $(g_{ab}, \nabla_a)$  is called the intrinsic  geometry of $H$.}
\medskip
    
The derivative $\nabla_a$ in $H$ is torsion free and satisfies the pseudo-metricity condition 
\begin{equation}\label{Dh}   \nabla_c g_{ab}\ =\ 0.  \end{equation}
It follows from those properties of $\nabla_a$ and from the degeneracy of $g_{ab}$ that  
for every vector field $\ell$ tangent to $H$ and such that (\ref{ll}), the Lie derivative of $g_{ab}$ vanishes \cite{Lewandowski:2004sh}
\begin{equation}\label{Llh}
{\cal L}_\ell g_{ab}\ =\ 0 .
\end{equation}
That property  can be used as an equivalent (and perhaps clearer) definition of NE-SF null surfaces.  It is also equivalent to the existence of an extension $t$ of the vector field $\ell$ to a neighborhood of $H$ in $M$, such that
\be\label{Ltg} {\cal L}_tg_{\mu\nu}{}_{|_H}=0.\ee  
That extension is not unique, an example can be constructed as follows. 
To define the second ingredient it is convenient to choose a function $v:H\rightarrow \mathbb{R}$ such that
\be\label{v} \ell^a\nabla_av=1 .\ee
Let $n^\mu$ be the spacetime vector field defined in a neighborhood of $H$ by the following two conditions:
\begin{itemize}
\item[$(i)$]  $n^\mu$ is orthogonal to the sections of $H$ such that $v={\rm const}$; 
\item[$(ii)$] $\nabla_n n =0$. 
\end{itemize}
Using $n^\mu$, we define $t$ as follows
\be t_{|_H}=\ell, \ \ \ \ {\cal L}_n t = 0. \ee
It is easy to  show, that  $t$ satisfies (\ref{Ltg}).  
 
Hence, the null NE-SF surfaces can be thought of as the Killing horizons to the $0$th order. 

The property $(\ref{Llh})$ is invariant with respect to rescaling of the vector field $\ell$ by any  function $f$, 
\be    {\cal L}_\ell g_{ab}=0\ \ \ \Rightarrow\ \ \  {\cal L}_{f\ell}g_{ab}=0 . \ee

The induced covariant derivative $\nabla_a$ is constrained by the assumed lack of torsion and (\ref{Dh}), however it is not uniquely determined by those conditions. The remaining data is the rotation $1$-form potential and the transversal expansion-shear tensor. The rotation  $1$-form potential $\omega_a$ is defined for every choice of a vector field $\ell$ orthogonal to $H$, namely
\be \label{omega}\nabla_a\ell^b = \omega_a\ell^b . \ee
The pullback $S_{AB}$ of the  tensor
\be\label{S}  S_{ab}:= -\nabla_b \nabla_a v, \ee
 onto a constancy surface   of $v$ in $H$, is referred to as the transversal expansion-shear tensor.  Notice, that we can inverse the order between introducing $v$ and its constance surfaces: given any space-like section of $H$ we can choose a solution $v$ to the equation (\ref{v}) that is constant on the section. Hence, $S_{AB}$ is a property of a spacelike sections of $H$, given the vector field $\ell$.   
 The Einstein equations (\ref{EE}) induce constraints on NE-SF horizon geometry \cite{Lewandowski:2004sh}. They can be solved uniquely  given  on $H$: $\ell$, its surface gravity
 \be\label{kappa} \kappa=\ell^a\omega_a, \ee
  a single section 
 \be v= v_0 , \ee 
 and the pullbacks $g_{AB}$, $\omega_A$ and $S_{AB}$, respectively, of the  tensors $g_{ab}$, $\omega_a$, $S_{ab}$.   That data is free in the sense, that  when we vary all possible Einstein vacua with cosmological constant and all possible NE-SF null surfaces, the data defined above takes all the possible (functional) values.
\medskip

\subsection{Extremal isolated horizons and the extremal Killing horizons to the $2$nd order}\label{eIH}  
Suppose that a null NE-SF surface  $H$ of intrinsic geometry $(g_{ab}, \nabla_a)$  admits a nowhere vanishing vector field $\ell$ orthogonal to $H$  such that  
\begin{equation}\label{LlD}
[{\cal L}_\ell, \nabla_{a}]\ =\ 0 . 
\end{equation}   
In other words, the flow of $\ell$ preserves all the intrinsic geometry $(g_{ab},\nabla_a)$.  Then, we say that $H$ is isolated horizon (IH).  
 
\noindent{\bf Definition 2}  {\it Whenever a null symmetry $\ell$ of an IH $H$ has identically zero self acceleration, that is if}
 \be \ell^a\nabla_a\ell=0, \ee
 {\it  then we say that $H$ is extremal IH.}   
  
 Notice, that   the condition (\ref{LlD})  is  not invariant with respect to rescaling  $\ell$ by arbitrary function. In fact, for a generic IH, only  
 $$f= f_0={\rm const},$$
 preserves  (\ref{LlD}).   
 
The properties (\ref{Llh},\ref{LlD})  are  equivalent to the existence of an extension $t$ of the vector field $\ell$ to a neighborhood of $H$ in $M$, such that
\be\label{Ltgnabla} {\cal L}_tg_{\mu\nu}{}_{|_H}=0, \ \ \ {\rm and}\ \ \  [{\cal L}_t, \nabla_{\mu}]{}_{|_H}=\ 0 .\ee  
An example  of $t$, given the vector field $\ell$, is provided by the conditions $(i)$ and $(ii)$ in Sec. \ref{NE-SF}. 

In that sense, (extremal) IHs  can be called  (extremal) Killing horizons to the $1$st order.   

The Einstein equations (\ref{EE}) combined with the properties of extremal IHs amount to  farther  constraints on  the data defined on a section of an extremal IH $H$ in Sec. \ref{NE-SF} and determining the  horizon geometry $(g_{ab}, \nabla_a)$. Given a spacelike section $\mathcal{S}$ of an extremal IH $H$,  the pullbacks $g_{AB}$ and $\omega_A$ of the degenerate geometry $g_{ab}$ and the rotation $1$-form potential $\omega_a$,  respectively, satisfy the following constraint
\be \label{nhg} \stackrel{(n-2)}{\nabla_{(A}}\omega_{B)}\ +\  \omega_{A}\omega_{B} \ -\ \frac{1}{2} \stackrel{(n-2)}{R_{AB}} \ +\ \frac{1}{n-2}\Lambda g_{AB} =0
 \ee   
 where $\stackrel{(n-2)}{\nabla_A}$ is just the torsion free and metric connection of $g_{AB}$ induced on the section of $H$, and $\stackrel{(n-2)}{R_{AB}}$ is its Ricci tensor. 
 
On the other hand, the transversal expansion-shear tensor $S_{AB}$ is free, it can be set arbitrarily. 

On IH $H$, given  the vector field $\ell$ up to a constant, not only the tensor $g_{AB}$, but also the $1$-form $\omega_A$ are independent of choice of a spacelike  section of $H$ (identifying the sections in the obvious way). That is due to (\ref{LlD}), and 
\be \ell^a\omega_a=0. \ee On the other hand, $S_{AB}$ depends on sections via transformations
\be v= v' + f, \ \ \ \ell^a\nabla_a f=0, \ee
 that is
 \be\label{Strans}  S'_{AB} = S_{AB} +  \stackrel{(n-2)}{\nabla}_A\stackrel{(n-2)}{\nabla}{}_B f - \frac{1}{2} \stackrel{(n-2)}{\nabla}_{(A} \omega_{B)}.\ee
 It is also sensitive on rescaling of the vector field $\ell$ with a constant
 \be\label{Strans'}   \ell'=a\ell, \ \ \ \ \ \ v'=\frac{1}{a}v, \ \ \ \ \ \ S'_{AB}=\frac{1}{a}S_{AB}.  \ee  

In this paper we assume that there exists   an extension $t$ of the vector field $\ell$ to a neighborhood of $H$ in $M$, such that in addition to (\ref{Ltgnabla}) the spacetime Riemann tensor is Lie dragged along the horizon,
\be\label{LtR} {\cal L}_t R_{\alpha\beta\mu\nu}{}_{|_H}=0,\ee  
that makes  $H$ an extremal Killing horizon to the $2$nd  order.    We will show, that this very condition implies another   constraint on the intrinsic geometry $(g_{ab}, \nabla_a)$ of $H$, namely two linear second order partial differential equations on the transversal expansion-shear tensor $S_{AB}$. 
Moreover,   given an IH $H$, a null vector field $\ell$ defined on and orthogonal to $H$ such that (\ref{LlD}), and its extension $t$ determined by the conditions $(i)$ and $(ii)$ of Sec. \ref{NE-SF}, either the vector field $t$ satisfies also (\ref{LtR}) or there is no other extension that would coincide with $\ell$ on $H$, and satisfy all the three conditions (\ref{Ltgnabla},\ref{LtR}). \newline
It follows from the condition \eqref{LtR} that the metric is at least three times differentiable class and thus our manifold needs to be at least four times differentiable. Structure induced on the horizon must be at least two times differentiable which leads to the demand that the embedding is at least three times differentiable. No further assumptions regarding smoothness are made throughout the rest of the paper. 
\section{Extremal Killing horizons to the second order: the $4$-dimensional case}  \label{4d}
In this section we consider a $3$-dimensional extremal IH $H$ in  vacuum spacetime  $(M,g_{\mu\nu})$ with cosmological constant $\Lambda$ (\ref{EE}). We assume, that $H$ is an extremal Killing horizon to $2$-nd order according to the definition formulated at the end of Sec. \ref{eIH}, and derive new constraint on the horizon $H$ geometry $(g_{ab}, \nabla_a)$ of $H$. The notation we use is compatible with  the Newman-Penrose formalism when our $3$-frames are extended to null $4$-frames. 

\subsection{A null frame adapted to $H$ and the Newman-Penrose formalism} \label{adaptedframeH} 
On an IH $H$ introduced in  Sec. \ref{eIH} we have already defined a tangent null vector field $\ell$ and an adapted coordinate $v$ (\ref{v}). 
We complete  it to a null tangent to $H$ $3$-frame  $(m^a,\bar{m}^a,\ell^a)$, and the dual co-frame $(\bar{m}_a,{m}_a,\nabla_av=:-n_a)$ such that 
\begin{equation} \label{[l,m]}
{\cal L}_\ell m^a = 0.
\end{equation}
In this frame the isolated horizon $H$ geometry $(g_{ab}, \nabla_a)$ and its ingredients defined in Sec. \ref{NE-SF} can be expressed as follows
\begin{align}
g_{ab}\ &=\ m_a\bar{m}_b + m_b\bar{m}_a \\
\nabla_a \ell^b\  &= \  \left( (\alpha + \bar \beta)m_a + (\bar \alpha +\beta)\bar m_a \right) \ell^b =:\omega_a\ell^b,\\  
m^b  \nabla_a \bar m_b\ & =\ -(\alpha - \bar \beta)m_a + (\bar \alpha - \beta) \bar m_a  = - \bar m^b  \nabla_a m_b,\\
  S_{ab}\  &=\ \lambda m_am_b + \bar{ \lambda} \bar{m}_a \bar{m}_b + \mu (m_a\bar m_b+ m_b\bar m_a) 
  -  n_a\omega_b  -  n_b\omega_a. \label{Sab_NP}
\end{align}
The complex valued functions $\alpha,\beta,\lambda$ and the real function $\mu$ are defined by the equations above, and they are constant along the null generators of the horizon $H$ due to (\ref{LlD}). Consistently with the Newman-Penrose formalism, our notation  distinguishes between the vector fields, and the operators they define,
\be \label{NP2}D:= \ell^a \partial_a, \ \ \ \ \ \ \ \delta:= m^a\partial_a .\ee
In particular, it follows from (\ref{LlD}) that
\be D \alpha  = D\beta  = D\lambda  = D \mu  = 0. \ee
The data induced on a slice  $v=v_0$ of $H$  at the end of Sec. \ref{NE-SF}  is
\begin{align}
g_{AB}\ &=\ m_A\bar{m}_B+ m_B\bar{m}_A \label{NPg}\\
 \omega_A &= \   (\alpha + \bar \beta)m_A + (\bar \alpha +\beta)\bar m_A  ,\label{NPo}\\  
m^B  \stackrel{(2)}\nabla_A \bar m_B\ & =\ -(\alpha - \bar \beta)m_A + (\bar \alpha - \beta) \bar m_A  = - \bar m^B  \stackrel{(2)}\nabla_A m_B,
\label{NPnabla}\\
  S_{AB}\  &=\ \lambda m_A {m}_B + \bar{ \lambda} \bar{m}_A \bar{m}_B  + \mu (m_A\bar m_B +m_B\bar m_A) \label{SAB_NP2}
 \end{align}

We extend the  $3$-frame  $(m^a,\bar{m}^a,\ell^a)$ and the dual  $3$-co-frame $(\bar m_a,{m}_a, -n_a)$  into a null $4$-frame $(e_1{}^\mu, e_2{}^\mu, e_3{}^\mu, e_4{}^\mu)$ defined on  $H$,  in the following way
\be   e_1= m,\ \  e_2=\bar m, \ \ e_4=\ell   \ee
while the missing vector $e_3$ we determine  by the condition that in the dual co-frame $(e^1{}_\mu,...,e^4{}_\mu)$, the pullback of $e^4$ to $H$ is $-n_a$,
\be e^4{}_a = -n_a . \ee
The vector field $e^4$ is related with the vector field $n^\mu$ defined by the conditions $(i),(ii)$ in Sec. \ref{NE-SF}, namely
\be  e_3{}^\mu = -n^\mu . \ee

\subsection{The Weyl tensor and the Bianchi identity}
The advantage of that frame is, that if $t$ is the vector field in a neighborhood of $H$ in $M$ that satisfies the condition (\ref{Ltgnabla}), then
\be  {\cal L}_t e^\mu{}_{|_H} =0 = {\cal L}_t e_\mu{}_{|_H}  . \ee
Therefore, the condition (\ref{LtR}) amounts to the condition on the components of the Riemann tensor,
\be D\stackrel{(4)}R_{\mu\nu\alpha\beta}=0 ,\ee
and taking into account the Einstein equations (\ref{EE}), it becomes  the condition on  the Newman-Penrose components Weyl tensor,  
\be\label{DPsi} D\Psi_I{}_{|_H}  =0, \ \ \ \ \ I=0,1,2,3,4 \ee
where
 \begin{equation}
\Psi_0\ :=\ \stackrel{(4)}C_{4141}, \ \ \ \ \ \Psi_1\ :=\  \stackrel{(4)}C_{4341}\ \ \ \ \ \Psi_2\ :=\ \stackrel{(4)}C_{4123}, \ \ \ \ \ 
\Psi_3\ :=\ \stackrel{(4)}C_{3432}, \ \ \ \ \ \Psi_4\ :=\   \stackrel{(4)}C_{3232} 
\end{equation}
of the Weyl tensor $\stackrel{(4)}C_{\alpha\beta\mu\nu}$.  
Now, one of the Bianchi identity 
\be \stackrel{(4)}\nabla_\alpha\stackrel{(4)}C{}^\alpha{}_{\beta\mu\nu}=0 \ee
implies
\be
0\ =\ D\Psi_4 - \bar\delta \Psi_3 +3\lambda\Psi_2- 2(2\pi+\alpha)\Psi_3 + 2\kappa \Psi_4 ,\ee
where $\kappa$ is the surface gravity (\ref{kappa}) that vanishes in the extremal case we are considering here and $\pi=\alpha+\bar\beta$ is a component of $\omega_A$ in null frame. 
Via (\ref{DPsi}) the identity  becomes
\be
0\label{B}\ =\  - \bar\delta \Psi_3 +3\lambda\Psi_2- 2(2\pi+\alpha)\Psi_3 .\ee
The Weyl tensor components $\Psi_2$ and $\Psi_3$ are  determined at every vacuum NE-SF null surface with the cosmological constant $\Lambda$ by  $g_{AB}, \omega_A$
and $S_{AB}$, namely
\begin{align} \label{eq:psi21}
\Psi_{2} &= \bar\delta \beta - \delta \alpha  + \alpha\bar\alpha + \beta\bar\beta - 2\alpha\beta  + \Lambda/6.\\
\Psi_{3} &= \bar\delta\mu - \delta\lambda + \mu(\alpha+\bar\beta)+ \lambda(\bar\alpha- 3\beta) .
\end{align}  

In the argument above, we assumed the existence of the vector field $t$ in a neighborhood of a NE-SF $H$ that makes $H$ an extremal Killing horizon to the second order. It is easy to see, that  the converse statement is true: given an IH  $H$, its null symmetry generator $\ell$   such that (\ref{LlD}), and the extension $t$ of $\ell$ defined by $(i),(ii)$ in Sec. \ref{NE-SF}, whenever (\ref{B}) is true, so is (\ref{LtR}). 
 
\subsection{Summary of the result and consequences}    \label{summary1}
\noindent {\bf Theorem 1.} {\it Suppose $H$ is a $3$-dimensional NE-SF null surface contained in a $4$-dimensional spacetime $(M,g)$ that satisfies the vacuum Einstein equations with cosmological constant $\Lambda$.  If $H$ is  an extremal  Killing horizon to the second order, then it satisfies the following conditions: $(i)$ it admits a vector field $\ell^a$ that makes it an extremal IH according to Def. 2.  $(ii)$ The intrinsic geometry $(g_{ab}, \nabla_a)$ (see Def. 1) and the vector field $\ell^a$ satisfy the following equations defined on an arbitrary space-like section ${\cal S}$ of $H$:}
\be \label{nhg4} \stackrel{(2)}\nabla_{(A}\omega_{B)}\ +\  \omega_{A}\omega_{B} \ -\ \frac{1}{2} \stackrel{(2)}R_{AB} \ +\ \frac{1}{2}\Lambda g_{AB} =0
 \ee   
\begin{align}
\label{second4} \begin{split}
0 &=\bar \delta \delta \lambda - \bar \delta\bar \delta \mu - 
\left(\alpha + \bar\beta \right)\bar\delta\mu - \mu \left(\bar \delta 
\alpha - \bar \delta \bar \beta\right) - \left( \bar \alpha - 3 \beta 
\right)\bar\delta\lambda -\lambda \left(\bar \delta \bar \alpha - 3
  \bar \delta \beta \right)\\
&+ 3 \lambda \Psi_2 -2\left(2 \pi +  \alpha \right) \left(\bar \delta \mu 
- \delta \lambda + \mu (\alpha + \bar \beta) + \lambda (\bar \alpha 
-3  \beta) \right) \end{split}
\end{align}
{\it   where $g_{AB}$ and $\omega_A$ is the pullback of the metric $g_{ab}$ and the rotation
   1-form potential $\omega_a$ (12), respectively, to ${\cal S}$,
   $\lambda$ and $\mu$  are the components in the null 2-frame  (\ref{Sab_NP}, \ref{SAB_NP2}) of the pullback  to ${\cal S}$ of the tensor $S_{ab}$ (\ref{S}),  the operator $\delta$ is defined in \eqref{NP2} and the functions $\alpha,\beta$ are defined by $g_{AB}$ and $\omega_A$ via (\ref{NPg}-\ref{NPnabla}).}
 \medskip
 
\noindent{\bf Remarks:} 
\begin{itemize} 
\item It follows from our derivation, that the resulting equations are invariant with respect to the transformations of $S_{AB}$ (\ref{Strans}, \ref{Strans'}).  Given $\omega_A$ and $g_{AB}$, tensors $S_{AB}$ and $S'_{AB}$ related with each other by  (\ref{Strans}, \ref{Strans'}) define isomorphic extremal isolated horizons.  
\item The equation \eqref{nhg4} can be also written in terms of the N-P coefficients  \cite{Ashtekar:2001jb}. On the other hand, the second equation will be written covariantly in the next section in the general $n$ dimensional spacetime case. 
\end{itemize}

\medskip

Our result leads to a complete characterization of the extremal Kerr horizon:
\medskip

\noindent{\bf Corollary 1.} $3$-dimensional extremal isolated horizon $H$, $g_{ab}$, $\nabla_a$, $\ell$  is embeddable into the extremal Kerr spacetime if and only if it satisfies all the following  conditions:
\begin{itemize}
\item[i)] $\ell \not=0$ at every point of $H$;
\item[ii)]  The null generators  of $H$ define the map $H\rightarrow S$ such that $S$ is diffeomorphic to $2$-sphere, the map admits  a global section $S\rightarrow H$,  and $H$ is diffeomorphic to $\mathbb{R}\times S$;
\item[iii)] ($H$, $g_{ab}$, $\nabla_a$, $\ell$) is axially symmetric; 
\item[iv)] The geometry $g_{ab}$, $\nabla_a$  satisfies the equations (\ref{nhg4},\ref{second4});    
\item[v)] $\mu(x)\not=0$ for every $x\in S$.  
\end{itemize}

The Corollary follows from the uniqueness of the  axisymmetric solutions of (\ref{nhg4})  \cite{Lewandowski:2002ua, Kunduri:2013ana} and the uniqueness modulo the transformations (\ref{Strans}, \ref{Strans'}) of the axisymmetric solutions of the equation (\ref{second4})  \cite{Li:2015wsa}, both with $\Lambda=0$. 

\section{Extremal Killing horizons to the second order: the $n>2$-dimensional case}   \label{main}
\subsection{Conventions and null Gau\ss ian coordinates adapted to $H$}\label{con}
In this section we will generalize constraints on $S_{AB}$ to the case of arbitrary dimension $n$. As before, Einstein equation holds. To simplify our notation we introduce $n$-dimensional cosmological constant
\be
\Lambda_n = \frac{2}{n-2} \Lambda.
\ee
As before we denote dimensionality of curvature tensors via the numbers above them. We still assume that the horizon is extremal (it means, $\kappa = 0$) stationary to the second order.
In the neighbourhood of the horizon we can introduce null Gau\ss ian coordinates $(v,r, x^A)$ in which metric tensor reads:
\begin{equation}
     g = 2 dv (\frac{1}{2} r^2 f dv + dr + 2 r h_A dx^A) + \gamma_{AB} dx^A dx^B  \label{gauge}
\end{equation}
and $l = \partial_v$ and the horizon is located at $r=0$ and $f$, $h_A$ and $\gamma_{AB}$ are smooth. The $r^2$ factor in front of $dv^2$ comes from the assumption of extremality. A choice of $v$ coordinate is equivalent to the choice of a spatial section $\mathcal{S}$ of $H$. Spacetime metric induces lower dimensional objects on $\mathcal{S}$:
\begin{align}
\begin{split}
    &g_{AB} = \gamma_{AB}{}_{|_{r=0}} \\
    &\omega_A = h_{A}{}_{|_{r=0}} \\ \label{dane_na_hor}
    &S_{AB} = -\frac{1}{2} \gamma_{AB,r}{}_{|_{r=0}}
    \end{split}
\end{align}
One can easily check that objects defined in such a manner are exactly $g_{AB}$, $\omega_A$ and $S_{AB}$ defined before. \\
Now we are prepared to derive constraint equations on $S_{AB}$.
\subsection{Summary of the main result} \label{summary2}
Let us start by writing out explicitly our assumptions and the result. \\
\noindent {\bf Theorem 2.} {\it Suppose $H$ is a $(n-1)$-dimensional NE-SF null surface contained in a $n$-dimensional spacetime $(M,g)$ that satisfies the vacuum Einstein equations with a cosmological constant $\Lambda$. If $H$ is  an extremal  Killing horizon to the second order, then it satisfies the following conditions: $(i)$ it admits a vector field $\ell^a$ that makes it an extremal IH according to Def. 2. $(ii)$ The intrinsic geometry $(g_{ab}, \nabla_a)$ (see Def. 1) and the vector field $\ell^a$ satisfy the following equations defined on an arbitrary space-like section ${\cal S}$ of $H$:}

\be \label{nhg_n} \stackrel{(n-2)}\nabla_{(A}\omega_{B)}\ +\  \omega_{A}\omega_{B} \ -\ \frac{1}{2} \stackrel{(n-2)}{R_{AB}} \ +\ \frac{1}{n-2}\Lambda g_{AB} =0
 \ee   
{\it and an equation \eqref{constraint}, where $g_{AB}$ and $\omega_A$ is the pullback of the metric $g_{ab}$ and the rotation $1$-form potential $\omega_a$ (\ref{omega}), respectively,  and $S_{AB}$ is the  transversal expansion-shear  tensor.} 
 \medskip
\subsection{Sketch of a proof}
In this section we shall derive constraints on $S_{AB}$. Let us start with an observation that the condition $\mathcal{L}_l  R_{rArB}|_H = 0$ is equivalent to 
\begin{equation}
    g_{AB,rrv}|_H = 0. \label{2ndorder}
\end{equation}
Such a term will be present while calculating $R_{AB,r}$ which can be evaluated using Einstein equation \eqref{EE}. Before that, we need to find a few different transversal derivatives of $g_{\mu \nu}$. 
Equation
\begin{equation} \Lambda_n = R_{rv}      \end{equation}
gives us 
\begin{equation}
    2f = 2 \Lambda_n + 4\omega^2 - 2\omega_C^{ \ \ ;C} \label{f},
\end{equation} 
where semicolon denotes covariant derivative in $n-2$ dimensions and $\omega^2 = \omega^A \omega_A$, and thus, a solution $(g_{AB}, \omega_A)$ to \eqref{nhge} automatically fixes $f$.
From $R_{rA} = 0$ we obtain 
\begin{equation} 4 h_{A,r} = 2\omega_A S - 4 \omega^C S_{AC} - 2 S_{;A} + 2 S_{AC}^{\ \ ;C} \label{hAr},   \end{equation}
where $S = q^{AB} S_{AB}$ is the trace of $S_{AB}$. \eqref{hAr} determines transversal derivative of $h_A$ entirely through the geometrical objects on $\mathcal{S}$. 
Eventually, equation
\begin{equation} R_{AB} = \Lambda_n g_{AB} \end{equation}
evaluated at the horizon does not introduce any new information besides \eqref{nhge}. However, taking its derivative upon $\partial_r$
\begin{equation} R_{AB,r} = - 2 \Lambda_n S_{AB}       \end{equation}
and combining it with (\ref{2ndorder}, \ref{f}, \ref{hAr}) gives us following linear equation:
\begin{align}
\begin{split}
    &0 = \\
    &\stackrel{(n-2)}{\Delta}S_{AB} - S_{;AB} - 2S_{(B}^{\ \ C}\stackrel{(n-2)}{R}_{A)C} + 2 S^{CD} \stackrel{(n-2)}{R}_{ACBD}\\
    &+4\omega^C S_{C(A;B)} + 6 \omega_{(A}S_{;B)} - 6 \omega^C S_{AB;C} - 4 \omega_{(A}S_{B)C}^{\ \ \ \ ;C} \\
    &+4 S_{C(A} \omega_{B)}^{\ \ ;C} - 4 \omega^C_{\ \ ;(B}S_{A)C} - 4 \omega_A \omega_B S + 4\omega^2 S_{AB}. \label{constraint}
\end{split}
\end{align}
\section{Discussion} \label{dis}
The main result of this paper is  Theorem $1$  (Sec. \ref{summary1}) and Theorem $2$ (Sec. \ref{summary2})  about extremal isolated horizons in $4$-dimensional and, respectively, $n$-dimensional  vacuum spacetimes.  Specifically, the equation  (\ref{second4}, \ref{constraint}), respectively, is a new constraint on  intrinsic structure  $(g_{ab},\nabla_a, \ell^a)$ of arbitrary IH $H$  that is necessary for the existence of an embedding of $H$ into a vacuum spacetime as an extremal Killing horizon to the second order.   The new constraint combined with the NHG equation (\ref{nhg4}, \ref{nhg_n}) allow to uniquely characterize those  IH geometries  that can be embedded into extremal Kerr or extremal Carter (Kerr-(anti) de Sitter) spacetime  (Corollary 1).                  
In fact, the equation \eqref{constraint} was first derived in \cite{Li:2015wsa} as a linearized, Einstein equation around the near horizon geometry. It is not coincidence -- since NHG limit in our coordinates corresponds to $v \mapsto \epsilon^{-1}v, r \mapsto \epsilon r$, after taking $\epsilon \rightarrow 0$ our $S_{AB}$ behaves as a transversal mode in the whole spacetime. Another difference is that in our approach the horizon was not a Killing one but only stationary to the second order. \\
This constraint  \eqref{constraint}  shall allow one to classify possible extremal horizons of a given NHG. $4$-dimensional problem was partially solved in \cite{Li:2015wsa} under assumption of axial symmetry. According to our knowledge, no non-zero solutions to \eqref{constraint} are known without this symmetry. \\
Several technical remarks are in order. 
One can easily notice that the right hand side of \eqref{constraint} is traceless and thus we have $\frac{(n-2)(n-1)}{2} - 1$ equations. On the other hand, there are $\frac{(n-2)(n-1)}{2}$ components of $S_{AB}$. However, our choice of the coordinate $v$ in \eqref{gauge} is ambiguous.  There are residual transformations $v \mapsto v + F(x^A)$ under which $g_{AB}$ and $\omega_A$ are invariant but $S_{AB}$ is not. Thus, one could  impose one additional gauge fixing condition on $S_{AB}$.  The fact that above constraints are homogeneous in $S_{AB}$ comes from yet another residual gauge transformation, namely rescaling $v$ and $r$ by inverse constant factors. Under such coordinate change, $g_{AB}$ and $\omega_A$ are invariant but $S_{AB}$ transforms linearly and so $\eqref{constraint}$ must transform covariantly upon it.
\begin{acknowledgments}
We thank James Lucietti for a useful piece of information concerning his and Carmen Li's earlier results. JL and AS were  supported by the Polish National Science Centre grant OPUS No. 2015/17/B/ST2/.
\end{acknowledgments}
\appendix
\section{Calculations} \label{ap}
In this appendix we shall demonstrate explicit calculations leading to the equation \eqref{constraint}. \\
All equations are at $H$ unless stated otherwise. Let us start with calculating $R_{rv}$ to obtain \eqref{f}:
\begin{align}
    \begin{split}
        &\Lambda_n = R_{rv} = g^{\alpha \beta} R_{\alpha r \beta v} = R_{vrrv} + g^{CD} R_{CrDv} \\
 &R_{vrrv} = \frac{1}{2} g_{vv,rr} + g_{\alpha \beta} \left(\Gamma^\alpha_{rr}\Gamma^{\beta}_{vv} - \Gamma^\alpha_{rv}\Gamma^\beta_{rv} \right)       \\
&\Gamma^\alpha_{\ rv} = \frac{1}{2} g^{\alpha \beta} g_{v\beta,r} = \omega^A \delta^\alpha_{\ A}\\ 
&R_{vrrv} = \frac{1}{2} g_{vv,rr} -  \omega^2 \\
    &R_{CrDv} = 
    \frac{1}{2} g_{Cv,rD}
    + g_{\alpha \beta} \left(\Gamma^\alpha_{\ rD} \Gamma^\beta_{\ Cv} - \Gamma^\alpha_{\ rv} \Gamma^\beta_{\ CD}
\right)
\\
    &\Gamma^\alpha_{\ rD} = \frac{1}{2} g^{\alpha \beta} g_{D\beta,r} =  \delta^{\alpha}{}_{r} \omega_D - g^{\alpha E} S_{ED} \\
        &\Gamma^\alpha_{\ Cv} = \frac{1}{2}g^{\alpha \beta} \left(
    g_{v\beta,C} - g_{Cv,\beta}
    \right) = - \omega_C \delta^{\alpha}{}_{v} \\
    &R_{CrDv} =  \omega_{C,D} -  \omega_C \omega_D -  \omega_A \Gamma^{A}_{CD} =  \omega_{C;D} -  \omega_C \omega_D \\
    &\Lambda_n = R_{rv} = \frac{1}{2}g_{vv,rr} -  \omega^2 +  \omega_C^{;C} -  \omega^2 \\
     &f = g_{vv,rr} =  \Lambda_n + 2\omega^2 - \omega_C^{\ \ ;C}.
    \end{split}
\end{align}
As the next step, we want to find $h_{A,r}$ and to this goal one needs to calculate $R_{rA}$ which is zero due to Einstein equations:
\begin{align}
    \begin{split}
        &0=R_{rA} = g^{\alpha \beta} R_{\alpha r \beta A} = R_{v r r A} + g^{CD} R_{CrDA} \\
         &R_{v r r A} = 
    \frac{1}{2} g_{Av,rr}
    + g_{\alpha \beta} \left(
    \Gamma^\alpha_{rr} \Gamma^\beta_{Av} - \Gamma^\alpha_{rv} \Gamma^\beta_{Ar}
    \right) =  \frac{1}{2} g_{Av,rr} - g_{\alpha \beta}\Gamma^\alpha_{rv} \Gamma^\beta_{Ar}  \\
        &R_{v r r A} =
    \frac{1}{2} g_{Av,rr} - h_C \Gamma^C_{Ar} = \frac{1}{2} g_{Av,rr} + \omega^C S_{CA} \\
    &R_{CrDA} = \frac{1}{2}
\left(
    g_{AC,rD} - g_{CD,rA}        
\right) + g_{\alpha \beta} \left( 
\Gamma^\alpha_{rD} \Gamma^\beta_{AC} - \Gamma^\alpha_{CD} \Gamma^\beta_{Ar}
\right) \\
&R_{CrDA} = S_{CD,A} - S_{AC,D} +  \omega_D S_{AC} -  \omega_A S_{CD} - S_{BD} \Gamma^B_{AC} + S_{AB} \Gamma^B_{CD} \\
& R_{CrDA} = S_{CD;A} - S_{AC;D} + \omega_D S_{AC} -  \omega_A S_{CD} \\
    &0 = R_{rA} = 
    \frac{1}{2} g_{Av,rr} + \omega^C S_{CA} + S_{;A} - S_{AC}^{\ \ ;C} +  \omega^C S_{AC} - \omega_A S\\
    &    X_A:=h_{A,r}|_H = g_{Av,rr}|_H = 2\omega_A S - 4\omega^C S_{AC} - 2S_{;A} + 2 S_{AC}^{\ \ \ ;C}.
    \end{split} \label{X}
\end{align}    
The last line corresponds to the equation \eqref{hAr}. \\
Finally, we can calculate $R_{AB}$. One actually needs to take $R_{AB,r}$ and so it is necessary to evaluate it in the neighbourhood of the horizon.
\begin{align}
    \begin{split}
        &R_{AB} = g^{CD} R_{CADB} + g^{Cr} R_{CArB} + g^{rC} R_{rACB} + R_{rAvB} + R_{vArB}
\\
    &R_{AB,r} = 2 S^{CD} R^{(n-2)}_{CADB} + g^{CD} R_{CADB,r} - h^C \left( 
    R_{rACB} + R_{CArB}
    \right) + \left(R_{rAvB} + R_{vArB} \right)_{,r}
    \end{split}
\end{align}
Let us calculate necessary ingredients step by step.
\begin{align}
    \begin{split}
    &R_{rACB} = -R_{ArCB} = - \left( 
    S_{AC;B} - S_{BA;C} +  h_C S_{AB} - h_B S_{AC}
    \right)
\\ &R_{CArB} =- R_{BrCA} = - \left(
    S_{BC;A} - S_{BA;C} +  h_C S_{AB} - h_A S_{BC}
\right) \\
&- h^C \left( 
    R_{rACB} + R_{CArB}
    \right) = 2h^C S_{C(A;B)} - 2 h^C S_{AB;C} + 2h^2 S_{AB} - 2h^C h_{(A}S_{B)C},
    \end{split}
\end{align}
where the first two lines comes from our previous calculation of $R_{CrDA}$ after permuting indices and convenient renaming. The next terms are transversal derivatives of the Riemann tensor:
\begin{align}
    \begin{split}
    &R_{rAvB} = \frac{1}{2}
    g_{Av,Br} + g_{\alpha \beta} \left(\Gamma^\alpha_{Av} \Gamma^\beta_{Br} - \Gamma^\alpha_{rv} \Gamma^\beta_{AB} 
    \right) \\
    &\left( R_{rAvB} + R_{vArB} \right)_{,r} = \\
     &= X_{(A,B)} + \left[g_{\alpha \beta} \left(
     \Gamma^\alpha_{Av} \Gamma^\beta_{Br} + \Gamma^\alpha_{Bv} \Gamma^\beta_{Ar}  - 2\Gamma^\alpha_{rv} \Gamma^\beta_{AB}
     \right) \right]_{,r}
     \\
     &= X_{(A;B)} - 
    4 h_{(A}S_{B)C}h^C -2
     h_{(A}X_{B)} + 2h^C \left( S_{AC;B} + S_{BC;A} - S_{AB;C} \right) - \\&2f S_{AB} - 2S_B^C h_{[C;A]} - 2S_{A}^C h_{[C;B]}
     + 4h^2 S_{AB}
    \end{split}
\end{align}
And so:
\begin{align}
\begin{split}
    &R_{AB} - g^{CD} R_{CADB,r} = \\
    &= X_{(A;B)} - 8 h_{(A}S_{B)C}h^C - 2 h_{(A}X_{B)} + 2h^C \left( S_{AC;B} + S_{BC;A} - S_{AB;C} \right) - g_{vv,rr} S_{AB} + \\
    &+ 4h^C S_{C(A;B)} - 4 h^C S_{AB;C} + 4h^2 S_{AB} - 4h^C h_{(A}S_{B)C} + S^{CD}R^{(n-2)}_{CADB} = \\
    &= X_{(A;B)} + S^{CD}R^{(n-2)}_{CADB} - 12h^C h_{(A}S_{B)C} - 2h_{(A}X_{B)} - 2( \Lambda_n - h_C^{;C}) S_{AB} \\&+ 8h^C S_{C(A;B)} - 6h^C S_{AB;C} + 2S_B^C h_{[A;C]} + 2 S_A^C h_{[B;C]}
    \end{split}
\end{align}
We are left with the task of calculating $R_{CADB,r}$:
\begin{align}
    \begin{split}
&R_{CADB,r} = \\
    &S_{CD,AB} + S_{AB,CD} - S_{BC, AD} - S_{AD,BC} \\
    &+\left(2 S_{AD} h_{(C;B)} + S_{CB}h_{(A;D)}-S_{AB}h_{(C;D)} - S_{CD} h_{(A;B)}\right)\\
    &+2S_{EF} \left(\Gamma^F_{AD} \Gamma^E_{BC} - \Gamma^E_{AB} \Gamma^F_{CD} \right)     
    \\
    &R_{CADB,r} = \\
    &S_{AB;CD} + S_{CD;AB} - S_{BC;AD} - S_{AD;CB} + S_A^E \stackrel{(n-2)}{R}_{ECDB} + S_C^E \stackrel{(n-2)}{R}_{EABD} \\
    &+2\left(S_{AD} h_{(C;B)} + S_{CB} h_{(A;D)} - S_{AB} h_{(C;D)} - S_{CD} h_{(A;B)} \right) \\
        &g^{CD} R_{CADB,r} = \\
    &\stackrel{(n-2)}{\Delta} S_{AB} + S_{;AB} - S_{BC;A}^{\ \ \ \ \ \ \ ;C} - S_{AC\ \ ;B}^{\ \ \ \ ;C} - S_A^{\ \ C}\stackrel{(n-2)}{R}_{CB} + S^{CD} \stackrel{(n-2)}{R}_{CABD} \\
    &+ 2\left(S_A^{\ \ C} h_{(C;B)} + S_B^{\ \ C} h_{(C;A)} - S_{AB} h_C^{\ \ ;C} - S h_{(A;B)} \right)
    \end{split}
\end{align}
Putting all those ingredients together (and, again using Einstein equations) we obtain
\begin{align}
\begin{split}
    0 &= \\
     &\Delta S_{AB} + S_{;AB} - S_{BC;A}^{\ \ \ \ \ \ \ ;C} - S_{AC\ \ ;B}^{\ \ \ \ ;C} - S_A^{\ \ C}R_{CB} - S^{CD} R_{CABD} \\ 
    &+ X_{(A;B)} - 12h^C h_{(A}S_{B)C} - 2h_{(A}X_{B)} \\&+ 8h^C S_{C(A;B)} - 6h^C S_{AB;C} + 2S_B^C h_{A;C} + 2S_A^C h_{B;C} - 2S h_{(A;B)},
    \end{split}
\end{align}
which (after inserting explicit form of $X_A$ obtained in \eqref{X}) reproduces \eqref{constraint}. One can notice that any $\Lambda$-dependence is only implicit, through $g_{AB}$ and $\omega_A$ which must satisfy \eqref{nhge}.
\bibliography{bibl.bib}

\begin{thebibliography}{19}%
\makeatletter
\providecommand \@ifxundefined [1]{%
 \@ifx{#1\undefined}
}%
\providecommand \@ifnum [1]{%
 \ifnum #1\expandafter \@firstoftwo
 \else \expandafter \@secondoftwo
 \fi
}%
\providecommand \@ifx [1]{%
 \ifx #1\expandafter \@firstoftwo
 \else \expandafter \@secondoftwo
 \fi
}%
\providecommand \natexlab [1]{#1}%
\providecommand \enquote  [1]{``#1''}%
\providecommand \bibnamefont  [1]{#1}%
\providecommand \bibfnamefont [1]{#1}%
\providecommand \citenamefont [1]{#1}%
\providecommand \href@noop [0]{\@secondoftwo}%
\providecommand \href [0]{\begingroup \@sanitize@url \@href}%
\providecommand \@href[1]{\@@startlink{#1}\@@href}%
\providecommand \@@href[1]{\endgroup#1\@@endlink}%
\providecommand \@sanitize@url [0]{\catcode `\\12\catcode `\$12\catcode
  `\&12\catcode `\#12\catcode `\^12\catcode `\_12\catcode `\%12\relax}%
\providecommand \@@startlink[1]{}%
\providecommand \@@endlink[0]{}%
\providecommand \url  [0]{\begingroup\@sanitize@url \@url }%
\providecommand \@url [1]{\endgroup\@href {#1}{\urlprefix }}%
\providecommand \urlprefix  [0]{URL }%
\providecommand \Eprint [0]{\href }%
\providecommand \doibase [0]{http://dx.doi.org/}%
\providecommand \selectlanguage [0]{\@gobble}%
\providecommand \bibinfo  [0]{\@secondoftwo}%
\providecommand \bibfield  [0]{\@secondoftwo}%
\providecommand \translation [1]{[#1]}%
\providecommand \BibitemOpen [0]{}%
\providecommand \bibitemStop [0]{}%
\providecommand \bibitemNoStop [0]{.\EOS\space}%
\providecommand \EOS [0]{\spacefactor3000\relax}%
\providecommand \BibitemShut  [1]{\csname bibitem#1\endcsname}%
\let\auto@bib@innerbib\@empty
\bibitem [{\citenamefont {Strominger}(1998)}]{Strominger:1997eq}%
  \BibitemOpen
  \bibfield  {author} {\bibinfo {author} {\bibfnamefont {A.}~\bibnamefont
  {Strominger}},\ }\href {\doibase 10.1088/1126-6708/1998/02/009} {\bibfield
  {journal} {\bibinfo  {journal} {JHEP}\ }\textbf {\bibinfo {volume} {02}},\
  \bibinfo {pages} {009} (\bibinfo {year} {1998})},\ \Eprint
  {http://arxiv.org/abs/hep-th/9712251} {arXiv:hep-th/9712251 [hep-th]}
  \BibitemShut {NoStop}%
\bibitem [{\citenamefont {Guica}\ \emph {et~al.}(2009)\citenamefont {Guica},
  \citenamefont {Hartman}, \citenamefont {Song},\ and\ \citenamefont
  {Strominger}}]{Guica:2008mu}%
  \BibitemOpen
  \bibfield  {author} {\bibinfo {author} {\bibfnamefont {M.}~\bibnamefont
  {Guica}}, \bibinfo {author} {\bibfnamefont {T.}~\bibnamefont {Hartman}},
  \bibinfo {author} {\bibfnamefont {W.}~\bibnamefont {Song}}, \ and\ \bibinfo
  {author} {\bibfnamefont {A.}~\bibnamefont {Strominger}},\ }\href {\doibase
  10.1103/PhysRevD.80.124008} {\bibfield  {journal} {\bibinfo  {journal} {Phys.
  Rev.}\ }\textbf {\bibinfo {volume} {D80}},\ \bibinfo {pages} {124008}
  (\bibinfo {year} {2009})},\ \Eprint {http://arxiv.org/abs/0809.4266}
  {arXiv:0809.4266 [hep-th]} \BibitemShut {NoStop}%
\bibitem [{\citenamefont {Amsel}\ \emph {et~al.}(2010)\citenamefont {Amsel},
  \citenamefont {Horowitz}, \citenamefont {Marolf},\ and\ \citenamefont
  {Roberts}}]{Amsel:2009et}%
  \BibitemOpen
  \bibfield  {author} {\bibinfo {author} {\bibfnamefont {A.~J.}\ \bibnamefont
  {Amsel}}, \bibinfo {author} {\bibfnamefont {G.~T.}\ \bibnamefont {Horowitz}},
  \bibinfo {author} {\bibfnamefont {D.}~\bibnamefont {Marolf}}, \ and\ \bibinfo
  {author} {\bibfnamefont {M.~M.}\ \bibnamefont {Roberts}},\ }\href {\doibase
  10.1103/PhysRevD.81.024033} {\bibfield  {journal} {\bibinfo  {journal} {Phys.
  Rev.}\ }\textbf {\bibinfo {volume} {D81}},\ \bibinfo {pages} {024033}
  (\bibinfo {year} {2010})},\ \Eprint {http://arxiv.org/abs/0906.2367}
  {arXiv:0906.2367 [gr-qc]} \BibitemShut {NoStop}%
\bibitem [{\citenamefont {Chrusciel}\ \emph {et~al.}(2012)\citenamefont
  {Chrusciel}, \citenamefont {Lopes~Costa},\ and\ \citenamefont
  {Heusler}}]{Chrusciel:2012jk}%
  \BibitemOpen
  \bibfield  {author} {\bibinfo {author} {\bibfnamefont {P.~T.}\ \bibnamefont
  {Chrusciel}}, \bibinfo {author} {\bibfnamefont {J.}~\bibnamefont
  {Lopes~Costa}}, \ and\ \bibinfo {author} {\bibfnamefont {M.}~\bibnamefont
  {Heusler}},\ }\href {\doibase 10.12942/lrr-2012-7} {\bibfield  {journal}
  {\bibinfo  {journal} {Living Rev. Rel.}\ }\textbf {\bibinfo {volume} {15}},\
  \bibinfo {pages} {7} (\bibinfo {year} {2012})},\ \Eprint
  {http://arxiv.org/abs/1205.6112} {arXiv:1205.6112 [gr-qc]} \BibitemShut
  {NoStop}%
\bibitem [{\citenamefont {Bardeen}\ and\ \citenamefont
  {Horowitz}(1999)}]{Bardeen:1999px}%
  \BibitemOpen
  \bibfield  {author} {\bibinfo {author} {\bibfnamefont {J.~M.}\ \bibnamefont
  {Bardeen}}\ and\ \bibinfo {author} {\bibfnamefont {G.~T.}\ \bibnamefont
  {Horowitz}},\ }\href {\doibase 10.1103/PhysRevD.60.104030} {\bibfield
  {journal} {\bibinfo  {journal} {Phys. Rev.}\ }\textbf {\bibinfo {volume}
  {D60}},\ \bibinfo {pages} {104030} (\bibinfo {year} {1999})},\ \Eprint
  {http://arxiv.org/abs/hep-th/9905099} {arXiv:hep-th/9905099 [hep-th]}
  \BibitemShut {NoStop}%
\bibitem [{\citenamefont {Pawlowski}\ \emph {et~al.}(2004)\citenamefont
  {Pawlowski}, \citenamefont {Lewandowski},\ and\ \citenamefont
  {Jezierski}}]{Pawlowski:2003ys}%
  \BibitemOpen
  \bibfield  {author} {\bibinfo {author} {\bibfnamefont {T.}~\bibnamefont
  {Pawlowski}}, \bibinfo {author} {\bibfnamefont {J.}~\bibnamefont
  {Lewandowski}}, \ and\ \bibinfo {author} {\bibfnamefont {J.}~\bibnamefont
  {Jezierski}},\ }\href {\doibase 10.1088/0264-9381/21/4/033} {\bibfield
  {journal} {\bibinfo  {journal} {Class. Quant. Grav.}\ }\textbf {\bibinfo
  {volume} {21}},\ \bibinfo {pages} {1237} (\bibinfo {year} {2004})},\ \Eprint
  {http://arxiv.org/abs/gr-qc/0306107} {arXiv:gr-qc/0306107 [gr-qc]}
  \BibitemShut {NoStop}%
\bibitem [{\citenamefont {Lewandowski}\ \emph {et~al.}(2016)\citenamefont
  {Lewandowski}, \citenamefont {Szereszewski},\ and\ \citenamefont
  {Waluk}}]{Lewandowski:2016sou}%
  \BibitemOpen
  \bibfield  {author} {\bibinfo {author} {\bibfnamefont {J.}~\bibnamefont
  {Lewandowski}}, \bibinfo {author} {\bibfnamefont {A.}~\bibnamefont
  {Szereszewski}}, \ and\ \bibinfo {author} {\bibfnamefont {P.}~\bibnamefont
  {Waluk}},\ }\href {\doibase 10.1103/PhysRevD.94.064018} {\bibfield  {journal}
  {\bibinfo  {journal} {Phys. Rev.}\ }\textbf {\bibinfo {volume} {D94}},\
  \bibinfo {pages} {064018} (\bibinfo {year} {2016})},\ \Eprint
  {http://arxiv.org/abs/1605.07038} {arXiv:1605.07038 [gr-qc]} \BibitemShut
  {NoStop}%
\bibitem [{\citenamefont {Hájiček}(1974)}]{Hajicek:1974oua}%
  \BibitemOpen
  \bibfield  {author} {\bibinfo {author} {\bibfnamefont {P.}~\bibnamefont
  {Hájiček}},\ }\href {\doibase 10.1007/BF01646202} {\bibfield  {journal}
  {\bibinfo  {journal} {Commun. Math. Phys.}\ }\textbf {\bibinfo {volume}
  {36}},\ \bibinfo {pages} {305} (\bibinfo {year} {1974})}\BibitemShut
  {NoStop}%
\bibitem [{\citenamefont {Moncrief}\ and\ \citenamefont
  {Isenberg}(1983)}]{Moncrief:1983xua}%
  \BibitemOpen
  \bibfield  {author} {\bibinfo {author} {\bibfnamefont {V.}~\bibnamefont
  {Moncrief}}\ and\ \bibinfo {author} {\bibfnamefont {J.}~\bibnamefont
  {Isenberg}},\ }\href {\doibase 10.1007/BF01214662} {\bibfield  {journal}
  {\bibinfo  {journal} {Commun. Math. Phys.}\ }\textbf {\bibinfo {volume}
  {89}},\ \bibinfo {pages} {387} (\bibinfo {year} {1983})}\BibitemShut
  {NoStop}%
\bibitem [{\citenamefont {Ashtekar}\ \emph {et~al.}(2002)\citenamefont
  {Ashtekar}, \citenamefont {Beetle},\ and\ \citenamefont
  {Lewandowski}}]{Ashtekar:2001jb}%
  \BibitemOpen
  \bibfield  {author} {\bibinfo {author} {\bibfnamefont {A.}~\bibnamefont
  {Ashtekar}}, \bibinfo {author} {\bibfnamefont {C.}~\bibnamefont {Beetle}}, \
  and\ \bibinfo {author} {\bibfnamefont {J.}~\bibnamefont {Lewandowski}},\
  }\href {\doibase 10.1088/0264-9381/19/6/311} {\bibfield  {journal} {\bibinfo
  {journal} {Class. Quant. Grav.}\ }\textbf {\bibinfo {volume} {19}},\ \bibinfo
  {pages} {1195} (\bibinfo {year} {2002})},\ \Eprint
  {http://arxiv.org/abs/gr-qc/0111067} {arXiv:gr-qc/0111067 [gr-qc]}
  \BibitemShut {NoStop}%
\bibitem [{\citenamefont {Lewandowski}\ and\ \citenamefont
  {Pawlowski}(2005)}]{Lewandowski:2004sh}%
  \BibitemOpen
  \bibfield  {author} {\bibinfo {author} {\bibfnamefont {J.}~\bibnamefont
  {Lewandowski}}\ and\ \bibinfo {author} {\bibfnamefont {T.}~\bibnamefont
  {Pawlowski}},\ }\href {\doibase 10.1088/0264-9381/22/9/007} {\bibfield
  {journal} {\bibinfo  {journal} {Class. Quant. Grav.}\ }\textbf {\bibinfo
  {volume} {22}},\ \bibinfo {pages} {1573} (\bibinfo {year} {2005})},\ \Eprint
  {http://arxiv.org/abs/gr-qc/0410146} {arXiv:gr-qc/0410146 [gr-qc]}
  \BibitemShut {NoStop}%
\bibitem [{\citenamefont {Lewandowski}\ and\ \citenamefont
  {Pawlowski}(2003)}]{Lewandowski:2002ua}%
  \BibitemOpen
  \bibfield  {author} {\bibinfo {author} {\bibfnamefont {J.}~\bibnamefont
  {Lewandowski}}\ and\ \bibinfo {author} {\bibfnamefont {T.}~\bibnamefont
  {Pawlowski}},\ }\href {\doibase 10.1088/0264-9381/20/4/303} {\bibfield
  {journal} {\bibinfo  {journal} {Class. Quant. Grav.}\ }\textbf {\bibinfo
  {volume} {20}},\ \bibinfo {pages} {587} (\bibinfo {year} {2003})},\ \Eprint
  {http://arxiv.org/abs/gr-qc/0208032} {arXiv:gr-qc/0208032 [gr-qc]}
  \BibitemShut {NoStop}%
\bibitem [{\citenamefont {Kunduri}\ and\ \citenamefont
  {Lucietti}(2009)}]{Kunduri:2008rs}%
  \BibitemOpen
  \bibfield  {author} {\bibinfo {author} {\bibfnamefont {H.~K.}\ \bibnamefont
  {Kunduri}}\ and\ \bibinfo {author} {\bibfnamefont {J.}~\bibnamefont
  {Lucietti}},\ }\href {\doibase 10.1063/1.3190480} {\bibfield  {journal}
  {\bibinfo  {journal} {J. Math. Phys.}\ }\textbf {\bibinfo {volume} {50}},\
  \bibinfo {pages} {082502} (\bibinfo {year} {2009})},\ \Eprint
  {http://arxiv.org/abs/0806.2051} {arXiv:0806.2051 [hep-th]} \BibitemShut
  {NoStop}%
\bibitem [{\citenamefont {Chruściel}\ \emph {et~al.}(2018)\citenamefont
  {Chruściel}, \citenamefont {Szybka},\ and\ \citenamefont
  {Tod}}]{Chrusciel:2017vie}%
  \BibitemOpen
  \bibfield  {author} {\bibinfo {author} {\bibfnamefont {P.~T.}\ \bibnamefont
  {Chruściel}}, \bibinfo {author} {\bibfnamefont {S.~J.}\ \bibnamefont
  {Szybka}}, \ and\ \bibinfo {author} {\bibfnamefont {P.}~\bibnamefont {Tod}},\
  }\href {\doibase 10.1088/1361-6382/aa90e7} {\bibfield  {journal} {\bibinfo
  {journal} {Class. Quant. Grav.}\ }\textbf {\bibinfo {volume} {35}},\ \bibinfo
  {pages} {015002} (\bibinfo {year} {2018})},\ \Eprint
  {http://arxiv.org/abs/1707.01118} {arXiv:1707.01118 [gr-qc]} \BibitemShut
  {NoStop}%
\bibitem [{\citenamefont {Chrusciel}\ \emph {et~al.}(2006)\citenamefont
  {Chrusciel}, \citenamefont {Reall},\ and\ \citenamefont
  {Tod}}]{Chrusciel:2005pa}%
  \BibitemOpen
  \bibfield  {author} {\bibinfo {author} {\bibfnamefont {P.~T.}\ \bibnamefont
  {Chrusciel}}, \bibinfo {author} {\bibfnamefont {H.~S.}\ \bibnamefont
  {Reall}}, \ and\ \bibinfo {author} {\bibfnamefont {P.}~\bibnamefont {Tod}},\
  }\href {\doibase 10.1088/0264-9381/23/2/018} {\bibfield  {journal} {\bibinfo
  {journal} {Class. Quant. Grav.}\ }\textbf {\bibinfo {volume} {23}},\ \bibinfo
  {pages} {549} (\bibinfo {year} {2006})},\ \Eprint
  {http://arxiv.org/abs/gr-qc/0512041} {arXiv:gr-qc/0512041 [gr-qc]}
  \BibitemShut {NoStop}%
\bibitem [{\citenamefont {Dobkowski-Ryłko}\ \emph {et~al.}(2018)\citenamefont
  {Dobkowski-Ryłko}, \citenamefont {Kamiński}, \citenamefont {Lewandowski},\
  and\ \citenamefont {Szereszewski}}]{Dobkowski-Rylko:2018nti}%
  \BibitemOpen
  \bibfield  {author} {\bibinfo {author} {\bibfnamefont {D.}~\bibnamefont
  {Dobkowski-Ryłko}}, \bibinfo {author} {\bibfnamefont {W.}~\bibnamefont
  {Kamiński}}, \bibinfo {author} {\bibfnamefont {J.}~\bibnamefont
  {Lewandowski}}, \ and\ \bibinfo {author} {\bibfnamefont {A.}~\bibnamefont
  {Szereszewski}},\ }\href {\doibase 10.1016/j.physletb.2018.08.048} {\bibfield
   {journal} {\bibinfo  {journal} {Phys. Lett.}\ }\textbf {\bibinfo {volume}
  {B785}},\ \bibinfo {pages} {381} (\bibinfo {year} {2018})},\ \Eprint
  {http://arxiv.org/abs/1807.05934} {arXiv:1807.05934 [gr-qc]} \BibitemShut
  {NoStop}%
\bibitem [{\citenamefont {Kunduri}\ and\ \citenamefont
  {Lucietti}(2013)}]{Kunduri:2013ana}%
  \BibitemOpen
  \bibfield  {author} {\bibinfo {author} {\bibfnamefont {H.~K.}\ \bibnamefont
  {Kunduri}}\ and\ \bibinfo {author} {\bibfnamefont {J.}~\bibnamefont
  {Lucietti}},\ }\href {\doibase 10.12942/lrr-2013-8} {\bibfield  {journal}
  {\bibinfo  {journal} {Living Rev. Rel.}\ }\textbf {\bibinfo {volume} {16}},\
  \bibinfo {pages} {8} (\bibinfo {year} {2013})},\ \Eprint
  {http://arxiv.org/abs/1306.2517} {arXiv:1306.2517 [hep-th]} \BibitemShut
  {NoStop}%
\bibitem [{\citenamefont {Wald}(1984)}]{Wald}%
  \BibitemOpen
  \bibfield  {author} {\bibinfo {author} {\bibfnamefont {R.~M.}\ \bibnamefont
  {Wald}},\ }\href {\doibase 10.7208/chicago/9780226870373.001.0001} {\emph
  {\bibinfo {title} {{General Relativity}}}}\ (\bibinfo  {publisher} {Chicago
  Univ. Pr.},\ \bibinfo {address} {Chicago, USA},\ \bibinfo {year}
  {1984})\BibitemShut {NoStop}%
\bibitem [{\citenamefont {Li}\ and\ \citenamefont
  {Lucietti}(2016)}]{Li:2015wsa}%
  \BibitemOpen
  \bibfield  {author} {\bibinfo {author} {\bibfnamefont {C.}~\bibnamefont
  {Li}}\ and\ \bibinfo {author} {\bibfnamefont {J.}~\bibnamefont {Lucietti}},\
  }\href {\doibase 10.1088/0264-9381/33/7/075015} {\bibfield  {journal}
  {\bibinfo  {journal} {Class. Quant. Grav.}\ }\textbf {\bibinfo {volume}
  {33}},\ \bibinfo {pages} {075015} (\bibinfo {year} {2016})},\ \Eprint
  {http://arxiv.org/abs/1509.03469} {arXiv:1509.03469 [gr-qc]} \BibitemShut
  {NoStop}%
\end{thebibliography}%
\end{document}